\documentclass[twocolumn]{aa}
\usepackage{graphicx}
\begin{document}

\authorrunning{Nagao et al.}
\titlerunning{Gas Metallicity of NLRs}

\title{Gas Metallicity in the Narrow-Line Regions of \\
       High-Redshift Active Galactic Nuclei}

\author{
          Tohru Nagao            \inst{1, 2},
          Roberto Maiolino       \inst{1}, \and
          Alessandro Marconi     \inst{1}
}

\offprints{T. Nagao}

\institute{
           INAF -- Osservatorio Astrofisico di Arcetri,
           Largo Enrico Fermi 5, 50125 Firenze, Italy\\
           \email{tohru@arcetri.astro.it,
                  maiolino@arcetri.astro.it,
                  marconi@arcetri.astro.it}
           \and
           National Astronomical Observatory of Japan,
           2-21-1 Osawa, Mitaka, Tokyo 151-8588, Japan
          }

\date{Received ;  accepted }

\abstract{
  We analyze optical (UV rest-frame) spectra of X-ray 
  selected narrow-line QSOs at redshift 
  $1.5 \la z \la 3.7$ found in the
  Chandra Deep Field South and of narrow-line radio 
  galaxies at redshift 
  $1.2 \la z \la 3.8$ to investigate the gas 
  metallicity of the narrow-line regions and their 
  evolution in this redshift range.
  Such spectra are also compared with UV spectra of 
  local Seyfert 2 galaxies. 
  The observational data are inconsistent with the
  predictions of shock models, suggesting that
  the narrow-line regions are mainly photoionized.
  The photoionization models with dust grains predict
  line flux ratios which are also in disagreement with most 
  of the observed values, suggesting that the high-ionization 
  part of the narrow-line regions (which is sampled by the
  available spectra) is dust-free.
  The photoionization dust-free models provide two
  possible scenarios which are consistent with the observed
  data: low-density gas clouds 
  ($n_{\rm H} \la 10^3$ cm$^{-3}$) with a sub-solar 
  metallicity 
  ($0.2 \la Z_{\rm gas}/Z_\odot \la 1.0$), or 
  high-density gas clouds
  ($n_{\rm H} \sim 10^5$ cm$^{-3}$) with a wide range of 
  gas metallicity
  ($0.2 \la Z_{\rm gas}/Z_\odot \la 5.0$).
  Regardless of the specific interpretation, the observational
  data do not show any evidence for a significant evolution 
  of the gas metallicity in the narrow-line regions
  within the redshift range $1.2 \la z \la 3.8$. Instead, we
  find a trend for more luminous active galactic 
  nuclei to have more 
  metal-rich gas clouds (luminosity-metallicity relation),
  which is in agreement with the same finding in the
  studies of the broad-line regions.
  The lack of evolution for the gas metallicity of the narrow-line
  regions implies that the major epoch of star formation in
  the host galaxies of these active galactic nuclei is at $z \ga 4$.
   \keywords{
             galaxies: active --
             galaxies: evolution --
             galaxies: nuclei --
             quasars: emission lines --
             quasars: general
            }
}

\maketitle

\section{Introduction}

Understanding galaxy formation and evolution is one of the
key astrophysical issues which is being pursued in this 
decade. The chemical composition of gas and stars in galaxies 
provides important information on this issue because it is 
a tracer of the star formation history in galaxies. A 
promising way to study the chemical evolution of galaxies 
is to measure their metallicity as a function of redshift. 
Since it is extremely hard and time-consuming to measure 
stellar metallicity of faint high-$z$ galaxies, because high 
quality spectra of shallow absorption features are required
(but see, e.g., Pettini et al. 2000; Mehlert et al. 2002), 
investigating the gas metallicity through emission lines is 
a promising strategy. However, most of the available diagnostic 
emission lines associated with massive star formation (e.g., 
[O{\sc ii}]$\lambda$3727, [O{\sc iii}]$\lambda\lambda$4959,5007,
[N{\sc ii}]$\lambda$6548,6583) are in rest-frame 
optical wavelength and thus are shifted into near infrared
in high-$z$ galaxies. Accordingly, observations 
of these emission lines in high-$z$ galaxies are feasible only 
for relatively bright targets (e.g., Teplitz et al. 2000; 
Pettini et al. 2001; Tecza et al. 2004; Shapley et al. 2004). 
Instead, active galactic nuclei (AGNs) exhibit bright 
emission lines at rest-frame UV wavelengths, which can be
used to investigate the gas metallicity even in high-$z$
objects. Spectroscopic observations of high-$z$ QSOs suggest 
that gas metallicity in the broad-line region (BLR) tends to be 
much higher than solar (e.g., Hamann \& Ferland 1992; 
Dietrich et al. 2003; Nagao et al. 2005) reaching as much as 
$Z \sim 15 Z_\odot$ (Baldwin et al. 2003). 
However it is not clear how the gas metallicity inferred from the
broad lines is related to the chemical properties of the host galaxies,
since the broad lines of AGNs sample only a very small 
region of galactic nuclei ($R_{\rm BLR} \ll$ 1 pc; e.g.,
Kaspi et al. 2000), which
may have evolved more rapidly than the host galaxy.

An alternative possibility is to use narrow line AGNs. In
this paper we focus on narrow-line AGNs at high redshift,
in particular on high-$z$ radio galaxies (HzRGs) and type-2 
QSOs (QSO2s). The nuclei of narrow-line AGNs are thought to
be obscured by edge-on optically thick tori (e.g., 
Antonucci \& Miller 1985; Antonucci 1993; Cohen et al. 1999). 
Since the broad emission lines and the strong ionizing 
continuum are blocked by this 
``natural coronagraph'', we can investigate narrow UV 
emission lines whose spatial extension is roughly comparable 
to that of the host galaxies ($\sim 10^{2-4}$ pc), without 
any complex deblending of broad and narrow components for 
the emission lines. Although narrow emission lines of 
HzRGs are often strongly influenced by a radio jet in terms 
of kinematics and morphological properties (e.g., McCarthy 
et al. 1991; Baum \& McCarthy 2000), the ionization 
mechanism is mostly dominated by photoionization, not by 
shock ionization (e.g., Villar-Martin et al. 1997; 
Allen et al. 1998; Iwamuro et al. 2003). 
Therefore, we can obtain information on the
gas metallicity of the narrow-line regions (NLRs) by comparing
the observed emission-line flux ratios with the predictions of
photoionization models.
By comparing the NLRs of HzRGs and QSO2s with low-$z$
type 2 AGNs (Seyfert 2 galaxies; Sy2s), we can investigate
whether the gas metallicity evolves on a cosmological
timescale or not.

By focusing on the flux ratio of 
N{\sc v}$\lambda$1240/C{\sc iv}$\lambda$1549 that is one of
the most frequently used metallicity diagnostics for AGNs
(generally for the BLR; e.g., Hamann \& Ferland 1992, 1999;
Dietrich et al. 2003), De Breuck et al. 
(2000) reported gas metallicity of HzRGs in the range
$0.4 Z_\odot \la Z_{\rm gas} \la 3.0 Z_\odot$ or 
possibly even much higher (see also van Ojik et al. 1994;
Vernet et al. 2001). They also claimed a metallicity evolution 
within their sample from $z > 3$ to $z < 3$:
the gas metallicity of all HzRGs at $z > 3$ in their sample is
$Z_{\rm gas} < 2 Z_\odot$, at variance with their sample 
at $z < 3$. Norman et al. (2002) reported that the strong 
N{\sc v}$\lambda$1240 emission of a X-ray selected 
high-$z$ QSO2 found in Chandra Deep Field South (CDFS;
Giacconi et al. 2002; Rosati et al. 2002), CDFS-202 
($z \simeq 3.7$), is consistent with a super-solar metallicity 
of its NLR, and more specifically $Z_{\rm gas} \sim 2.5-3.0 Z_\odot$.
However, since the N{\sc v}$\lambda$1240 emission becomes very
weak for metal-poor gas (i.e., $Z_{\rm gas} \la 1 Z_\odot$),
only upper-limit fluxes on N{\sc v}$\lambda$1240 are available
for the majority of the HzRG sample in De Breuck et al. (2000),
which makes the investigation of the metallicity evolution
difficult. An additional issue is that Iwamuro et al. (2003),
based on rest-frame optical spectra obtained by
sensitive near-infrared spectroscopy, 
recently reported that HzRGs at $2.0 < z < 2.6$ tend to show
sub-solar metallicities ($Z_{\rm gas} \sim 0.2 Z_\odot$),
significantly lower than values reported by
De Breuck et al. (2000). Although it is clear that improved and
additional observational data for a larger 
sample of HzRGs are required to reconcile this disagreement, 
both more sensitive measurements of N{\sc v}$\lambda$1240
and deeper near-infrared spectroscopic data are
very tough to obtain for HzRGs and QSO2s.

\begin{figure}
\resizebox{\hsize}{!}{\includegraphics{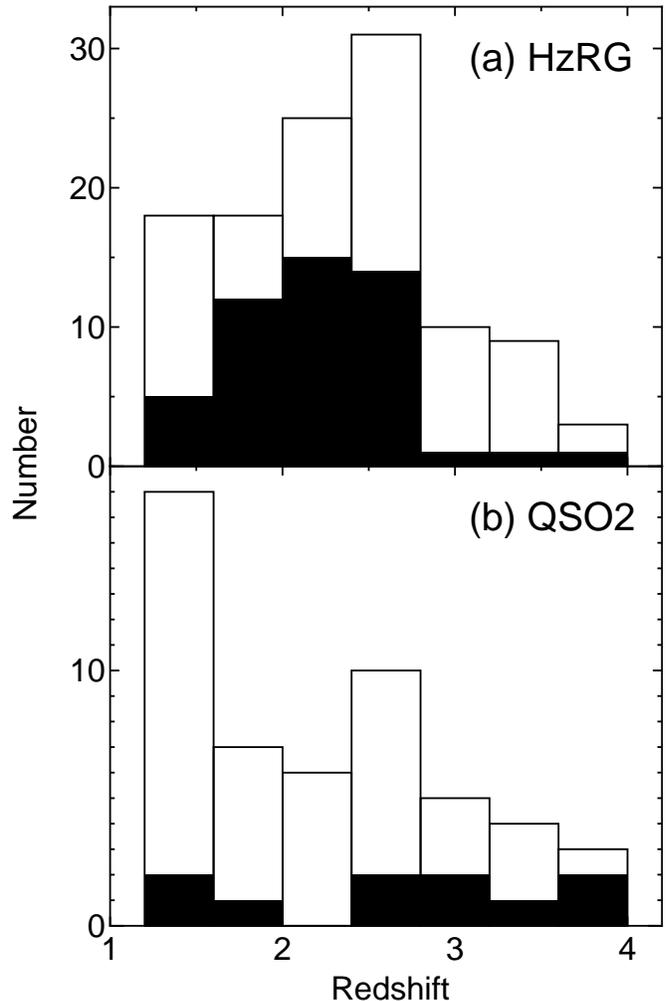}}
\caption{
   Redshift distributions of the sample objects.
   (a) The HzRG sample. The white histogram denotes 
   the original HzRG sample (at $1.2 \leq z \leq 4.0$;
   114 objects) of De Breuck et al. (2000) while the black 
   histogram denotes the 49 HzRGs whose emission-line fluxes 
   of C{\sc iv}$\lambda$1549, He{\sc ii}$\lambda$1640 and 
   C{\sc iii}]$\lambda$1909 are measured.
   (b) The QSO2 sample. The white histogram denotes
   the AGNs (including both broad-line and narrow-line AGNs)
   at $z \geq 1.2$ found in the CDFS (51 objects;
   Szokoly et al. 2004) and one additional QSO2 reported by 
   Stern et al. (2002), while the black histogram denotes the
   narrow-line AGNs among them (10 objects) which are 
   presented in Table 1.
}
\label{fig01}
\end{figure}

To tackle these issues, it is useful to find alternative 
metallicity diagnostics that consists only of strong
UV emission lines. In this paper, we investigate 
strong UV emission lines for HzRGs, QSO2s, and Sy2s.
By comparing the compiled data with photoionization model
calculations, we discuss the evolution of the gas metallicity 
of the NLRs in AGNs, from high-$z$ to the local universe,
using only strong UV emission lines. The compiled data and 
our photoionization model calculations are presented in 
\S2 and \S3, respectively. We compare the observational data with
the model results, discuss the properties of NLR gas clouds
and the implications for the chemical evolution of AGNs in \S4.
A brief summary is given in \S5.

%-----------------------------------------------------
%    Table 1
%-----------------------------------------------------

\begin{table*}
\centering
\caption{Emission-Line Measurements of X-ray selected QSO2s at $z > 1.2$}
\label{table:1}
\begin{tabular}{l c c c c c c}
\hline\hline
            \noalign{\smallskip}
  Object &
  Redshift &
  Ly$\alpha$$\lambda$1216$^{\mathrm{a}}$ & 
  N{\sc v}$\lambda$1240$^{\mathrm{a}}$ & 
  C{\sc iv}$\lambda$1549$^{\mathrm{a}}$ & 
  He{\sc ii}$\lambda$1640$^{\mathrm{a}}$ & 
  C{\sc iii}]$\lambda$1909$^{\mathrm{a}}$ \\
            \noalign{\smallskip}
\hline
            \noalign{\smallskip}
CDFS--027  & 3.064 & 
  12.6$\pm$0.7 & 
   2.5$\pm$0.7 & 
   6.4$\pm$0.5 &
   2.3$\pm$0.9 &
        $<$2.9$^{\mathrm{d}}$ \\
CDFS--031  & 1.603 & 
  ---          &
  ---          &
  24.1$\pm$1.4 &
  13.3$\pm$1.2 &
  10.3$\pm$1.3 \\
CDFS--057  & 2.562 & 
 112.2$\pm$1.3 &
   8.4$\pm$1.4 &
  17.8$\pm$0.8 &
   7.6$\pm$0.8 &
  13.3$\pm$0.9 \\
CDFS--112a$^{\mathrm{b}}$ & 2.940 & 
 175$\pm$6 &
  22$\pm$6 &
  30$\pm$6 &
  19$\pm$6 &
  15$\pm$6 \\
CDFS--153  & 1.536 &
  ---          &
  ---          &
  25.5$\pm$1.4 &
   6.2$\pm$1.1 &
  13.7$\pm$1.6 \\
CDFS--202  & 3.700 & 
  78.1$\pm$1.0 &
  26.8$\pm$1.1 &
  38.9$\pm$1.2 &
  19.7$\pm$1.5 &
       $<$12.9$^{\mathrm{d}}$ \\
CDFS--263b & 3.660 & 
  70.9$\pm$0.7 &
   4.6$\pm$0.7 &
  15.5$\pm$0.8 &
        $<$4.0$^{\mathrm{d}}$ &
        $<$7.6$^{\mathrm{d}}$ \\
CDFS--531  & 1.544 & 
  ---          &
  ---          &
  22.0$\pm$1.4 &
  17.4$\pm$1.5 &
  14.4$\pm$1.5 \\
CDFS--901  & 2.578 & 
  37.1$\pm$0.6 &
   6.5$\pm$0.8 &
  19.7$\pm$1.0 &
        $<$2.8$^{\mathrm{d}}$ &
   3.3$\pm$0.9 \\
CXO 52$^{\mathrm{c}}$ & 3.288 &
  189$\pm$4    &
    6$\pm$1.2$^{\mathrm{e}}$ &
   35$\pm$2    &
   17$\pm$2    &
   21$\pm$2    \\
            \noalign{\smallskip}
\hline
\end{tabular}
\begin{list}{}{}
\item[$^{\mathrm{a}}$]  
  Fluxes are in units of 10$^{-18}$ ergs s$^{-1}$ cm$^{-2}$.
\item[$^{\mathrm{b}}$]  
  FITS file not available for this source at the 
  CDFS web site at the time of writing of this
  paper; the fluxes and
  their errors were estimated from GIF file
  on the web site.
\item[$^{\mathrm{c}}$]  
  Data taken from Stern et al. (2002).
\item[$^{\mathrm{d}}$]  
  3$\sigma$ upper-limit flux.
\item[$^{\mathrm{e}}$]  
  The error was assigned by us; see text.
\end{list}
\end{table*}

%-----------------------------------------------------

\section{Data}

\subsection{Data compilation}

To investigate the possible chemical evolution of the NLRs 
in AGNs, we compiled the fluxes of some strong UV emission 
lines, N{\sc v}$\lambda$1240, C{\sc iv}$\lambda$1549, 
He{\sc ii}$\lambda$1640 and C{\sc iii}]$\lambda$1909.
We focus only on these emission lines since 
fainter lines were measured only in a few type 2 AGNs.
The details of the data compilation for each population 
of objects are given below.

(1) HzRGs: We used emission-line fluxes of HzRGs given by 
De Breuck et al. (2000), who compiled emission-line flux data 
of 165 radio galaxies from the literature. 
This sample contains 114 HzRGs at $1.2 \leq z \leq 3.8$
for which we can investigate the rest-frame UV spectra.
% In addition to them, upper-limit or lower-limit fluxes for 
% some additional HzRGs are also compiled.
We assigned the error 
on each line flux by checking the original references.
For objects whose emission-line flux errors are not given in the
original reference, we assigned a 20\% error for each emission line.
For emission-line fluxes measured with a statistical significance
lower than 3$\sigma$, we adopted 3$\sigma$ upper-limit values 
instead of the measured fluxes. 
We also referred to the measurements of emission-line flux data of 
9 HzRGs presented by Vernet et al. (2001). Although these HzRGs 
are included in the sample of De Breuck et al. (2000), we 
adopted the data of Vernet et al. (2001) of these 9 HzRGs
because of the higher signal-to-noise ratios of the 
Vernet et al. (2001) data.
Accordingly, the number of HzRGs whose fluxes of
C{\sc iv}$\lambda$1549, He{\sc ii}$\lambda$1640 
and C{\sc iii}]$\lambda$1909 are measured is 
% 49 (excluding upper-limit and lower-limit data).
51 in total, of which two objects are lower- or upper-limit data.

(2) QSO2s: We used the spectra of X-ray selected QSO2s
in CDFS recently obtained by Szokoly et al. (2004). 
Among 288 objects whose spectra
were obtained with VLT/FORS by Szokoly et al. (2004), 
51 objects are at $z > 1.2$. From these 51 objects,
we selected 9 objects with a high quality spectrum that shows
only narrow emission lines.
All of these 9 objects (distributed at $1.54 \la z \la 3.70$;
see Table 1) show strong C{\sc iv}$\lambda$1549
emission which indicates, along with their X-ray emission,
that these objects harbor an AGN
(i.e., not starburst galaxies). We obtained spectral data
of these 9 objects from the CDFS web 
site\footnote{http://www.mpe.mpg.de/CDFS/data/}, 
and measured their emission-line fluxes and errors
by means of a simple Gaussian fitting. The measured fluxes are given in 
Table 1. In this table fluxes of Ly$\alpha$ are also given 
for the 
reader's convenience, although we do not use the Ly$\alpha$ 
flux in the analysis and discussion in this paper. The reported
errors do not take any possible systematic errors into account.
In addition to the nine QSO2s in the CDFS, we also used
the emission-line flux data of CXO J084837.9+445352 (CXO 52),
a QSO2 at $z=3.288$ found by Stern et al. (2002).
Since the flux error for N{\sc v}$\lambda$1240 is
not given by Stern et al. (2002), we assigned a 20\% error 
for this line as for the HzRGs mentioned above.
Therefore the number of QSO2s is 10 in total.

(3) Sy2s: We compiled the flux data of 9 Sy2s observed by 
IUE from the literature. The object name, the compiled
flux data and their errors, and the reference are given in 
Table 2. For what concerns the data obtained by Thuan (1984),
note that they reported UV emission-line fluxes also for 
IC 5135, not only for NCG 5135; however, we do not include 
the data of IC 5135 because the signal-to-noise ratio of 
the IC 5135 data is too low (see Thuan 1984 for details).
For fluxes with no information on their errors in the
reference, we assigned a 20\% error as for the
HzRGs and QSO2s (except for low-accuracy measurements
explicitly mentioned in the reference for which we assigned a 
30\% error).
All of the targets are at $z < 0.03$ and thus we regard this
sample as representative of the NLRs in the local universe.

%-----------------------------------------------------
%    Table 2
%-----------------------------------------------------

\begin{table*}
\centering
\caption{Emission-Line Measurements of Sy2s}
\label{table:2}
\begin{tabular}{l c c c c c c c}
\hline\hline
            \noalign{\smallskip}
  Object &
  Redshift &
  Ly$\alpha$$\lambda$1216$^{\mathrm{a}}$ & 
  N{\sc v}$\lambda$1240$^{\mathrm{a}}$ & 
  C{\sc iv}$\lambda$1549$^{\mathrm{a}}$ & 
  He{\sc ii}$\lambda$1640$^{\mathrm{a}}$ & 
  C{\sc iii}]$\lambda$1909$^{\mathrm{a}}$ &
  Ref.$^{\mathrm{b}}$ \\
            \noalign{\smallskip}
\hline
            \noalign{\smallskip}
NGC 1068 & 0.004 &
  713$\pm$100 &
  224$\pm$41 &
  520$\pm$80 &
  187$\pm$29 &
  240$\pm$35 & 1 \\
NGC 4507$^{\mathrm{c}}$ & 0.012 &
  75.6$\pm$15.1 &
   5.2$\pm$1.0 &
  13.5$\pm$2.7 &
   5.6$\pm$1.1 &
   5.8$\pm$1.2 & 2 \\
NGC 5135$^{\mathrm{c}}$ & 0.014 &
  59.0$\pm$11.8 &
   1.1$\pm$0.2 &
   4.1$\pm$0.8 &
  10.0$\pm$2.0 &
  --- & 3 \\
NGC 5506$^{\mathrm{c}}$ & 0.006 &
  --- &
  --- &
  4.5$\pm$1.4$^{\mathrm{d}}$ &
  2.0$\pm$0.6$^{\mathrm{d}}$ &
  3.6$\pm$0.7 & 2 \\
NGC 7674 & 0.029 &
  47.0$\pm$20.3  &
  ---            &
  11.4$\pm$3.3   &
   5.1$\pm$1.5   &
   7.9$\pm$2.7   & 4 \\
Mrk 3    & 0.014 &
  64$\pm$19$^{\mathrm{d}}$ & 
   3$\pm$1$^{\mathrm{d}}$ &
  21$\pm$2 &
   9$\pm$1 &
   9$\pm$1 & 5 \\
Mrk 573 & 0.017 &
  151.2$\pm$22.7 & 
    6.3$\pm$0.9 & 
   29.0$\pm$4.3 & 
   12.6$\pm$1.9 & 
    8.8$\pm$1.3 &  6 \\
Mrk 1388 & 0.021 &
  --- & 
  --- & 
  8.3$\pm$1.2 & 
  3.8$\pm$0.6 & 
  3.6$\pm$0.5 & 6 \\
MCG --3--34--64$^{\mathrm{c}}$ & 0.017 &
  56$\pm$11 & 
   5$\pm$1 & 
  14$\pm$3 & 
  10$\pm$2 & 
   7$\pm$1 & 7 \\
            \noalign{\smallskip}
\hline
\end{tabular}
\begin{list}{}{}
\item[$^{\mathrm{a}}$]  
  Fluxes are in units of 10$^{-14}$ ergs s$^{-1}$ cm$^{-2}$.
\item[$^{\mathrm{b}}$]  
  References. ---
  (1) Snijders et al. 1986,
  (2) Bergeron et al. 1981,
  (3) Thuan 1984,
  (4) Kraemer et al. 1994,
  (5) Malkan \& Oke 1983,
  (6) McAlpine 1988,
  (7) De Robertis et al. 1988.
\item[$^{\mathrm{c}}$]  
  20\% of the line flux is assigned as the flux error.
\item[$^{\mathrm{d}}$]  
  30\% of the line flux is assigned as the flux error.
\end{list}
\end{table*}

The redshift distributions of the HzRG and the QSO2 samples
are shown in Figure 1.
The compiled data are not corrected for Galactic and
intrinsic dust reddening. Possible effects of the reddening
on our analysis are discussed later (\S\S2.2).
Since most of these data were obtained in low-dispersion
spectroscopy, the measured C{\sc iii}] flux may be
contaminated by the Si{\sc iii}]$\lambda\lambda$1883,1892 flux.
However, the contribution of Si{\sc iii}] is thought to be
small and we will discuss this issue further in \S\S3.2.

\subsection{Compilation results}

In Figure 2a the compiled line flux ratios are 
plotted on the
C{\sc iv}$\lambda$1549/He{\sc ii}$\lambda$1640 versus
C{\sc iii}]$\lambda$1909/C{\sc iv}$\lambda$1549 diagram.
The C{\sc iv}$\lambda$1549/He{\sc ii}$\lambda$1640 flux ratio
is expected to be sensitive to the gas metallicity.
This is because the gas temperature decreases when the 
metallicity increase in low-density 
($\la 10^4$ cm$^{-3}$) gas clouds 
(e.g., Ferland et al. 1984) and thus the collisional 
excitation of C{\sc iv} is gradually suppressed, while the 
He{\sc ii}$\lambda$1640 luminosity is basically proportional 
to the volume of the doubly-ionized helium region and thus 
rather insensitive to the gas metallicity.
The C{\sc iii}]$\lambda$1909/C{\sc iv}$\lambda$1549 ratio is
instead expected to be sensitive to the ionization degree of 
gas clouds. Therefore,
C{\sc iii}]$\lambda$1909/C{\sc iv}$\lambda$1549 can be used
to check any dependence of 
C{\sc iv}$\lambda$1549/He{\sc ii}$\lambda$1640 on the
ionization state of the gas. Summarizing, a diagnostic 
diagram that consists of these two flux ratios is expected to
be useful to estimate the properties of 
NLRs only with strong UV emission lines, as discussed more 
extensively in \S3 and \S4 (see also Groves et al. 2004).

In Figure 2 we also show the effect of the dust extinction 
on the line ratios, for the case of
$A_V = 5.0$ mag. An extinction curve described by
Cardelli et al. (1989) is adopted.
Since the dust extinction in typical
type 2 AGNs is generally $A_V \la 3$ mag
(see, e.g., Figure 5 of Nagao et al. 2001b), we conclude
that the effect of dust extinction on our analysis and
discussion is not important.

As clearly shown in Figure 2a, there is no significant 
difference in these two flux ratios between the high-$z$ QSO2 
sample and the local Sy2 sample. Some HzRGs show similar flux 
ratios to QSO2s and Sy2s, although other HzRGs show lower 
C{\sc iv}$\lambda$1549/He{\sc ii}$\lambda$1640 and higher
C{\sc iii}]$\lambda$1909/C{\sc iv}$\lambda$1549 than 
QSO2s and Sy2s. The logarithmically averaged ratios 
for these three populations 
(excluding upper-limits and lower-limits) are 
summarized in Table 3. 
These averaged flux ratios are also plotted in Figure 2b.
The average flux ratios of HzRGs appears systematically
different from those of QSO2s and Sy2s.
To see the statistical significance of the differences in
the line flux ratios between HzRGs and QSO2s -- Sy2s,
the Kolmogorov-Smirnov (K-S) statistical test is applied to
the data, discarding upper-limits and lower-limits.
The null hypothesis is that the flux ratios
(C{\sc iv}$\lambda$1549/He{\sc ii}$\lambda$1640 and
C{\sc iii}]$\lambda$1909/C{\sc iv}$\lambda$1549) of HzRGs
($N_{\rm data} = 49$) and QSO2s -- Sy2s ($N_{\rm data} = 14$)
come from the same underlying population. The derived
K-S probabilities are $4.0 \times 10^{-3}$ for
C{\sc iv}$\lambda$1549/He{\sc ii}$\lambda$1640 and
$8.0 \times 10^{-2}$ for
C{\sc iii}]$\lambda$1909/C{\sc iv}$\lambda$1549. These
results suggest that the difference in the 
C{\sc iv}$\lambda$1549/He{\sc ii}$\lambda$1640 ratio is
statistically significant while the difference in the
C{\sc iii}]$\lambda$1909/C{\sc iv}$\lambda$1549 ratio is
statistically marginal.

\begin{figure}
\centering
\rotatebox{-90}{\includegraphics[width=14.5cm]{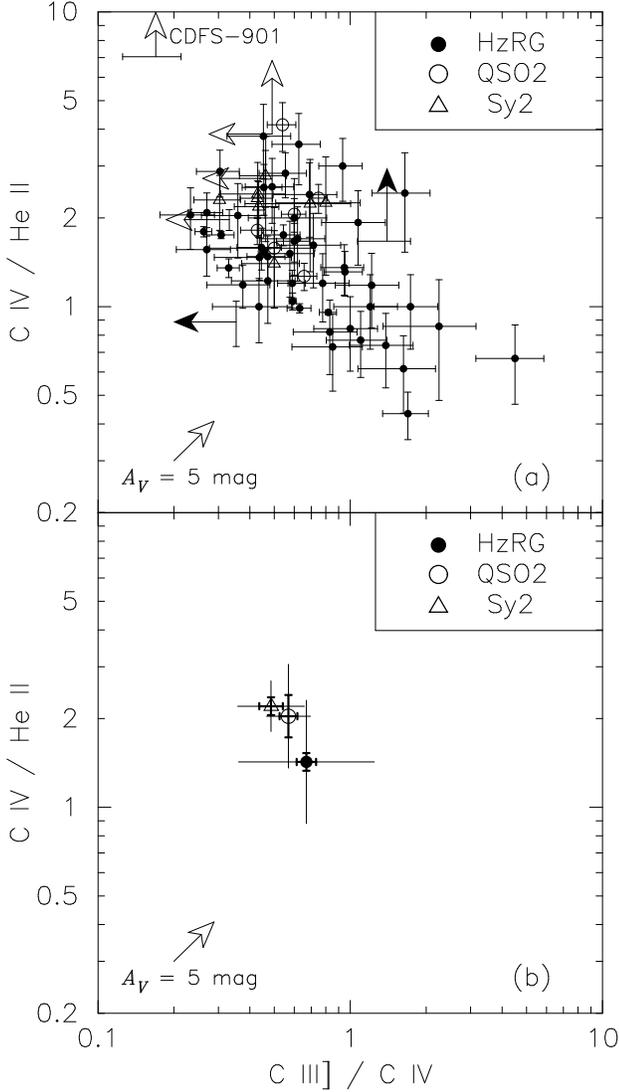}}
\caption{
   (a) Compiled data plotted on a diagram of 
   C{\sc iv}$\lambda$1549/He{\sc ii}$\lambda$1640 versus
   N{\sc v}$\lambda$1240/C{\sc iv}$\lambda$1549.
   Arrows denote the upper-limit or lower-limit on the flux ratios.
   Filled circles and arrows denote the HzRG data
   and open circles and arrows denote the QSO2 data.
   Open triangles denote the Sy2 data.
   The arrow at the lower-left corner in the panel denote
   the effect of a correction for dust extinction ($A_V = 5$ mag).
   (b) Logarithmically averaged flux ratios of
   HzRGs (filled circle), QSO2s (open circle) and Sy2s (open 
   triangle) plotted on a diagram of 
   C{\sc iv}$\lambda$1549/He{\sc ii}$\lambda$1640 versus
   N{\sc v}$\lambda$1240/C{\sc iv}$\lambda$1549.
   The arrow at the lower-left corner is the same as in (a).
   Thin bars denote the RMS's of the data distribution, and
   thick bars denote the estimated errors on the averaged values.
}
\label{fig02}
\end{figure}

When focusing only on HzRGs, an anti-correlation between
the emission-line flux ratios of 
C{\sc iv}$\lambda$1549/He{\sc ii}$\lambda$1640 and
C{\sc iii}]$\lambda$1909/C{\sc iv}$\lambda$1549 is apparent
in Figure 2a. To examine the statistical significance
of this correlation, the Spearman rank-order test is applied
to the HzRG data, discarding upper-limits and lower-limits.
The derived Spearman rank-order correlation
coefficient ($r_{\rm S}$) and its statistical significance
$p(r_{\rm S})$, which is the probability of the data being
consistent with the null hypothesis that the flux ratios are
not correlated, are 
$r_{\rm S} = -0.54$ and $p(r_{\rm S}) = 2.1 \times 10^{-4}$
($N_{\rm data} = 49$).
This indicates that the apparent anti-correlation between
C{\sc iv}$\lambda$1549/He{\sc ii}$\lambda$1640 and
C{\sc iii}]$\lambda$1909/C{\sc iv}$\lambda$1549 of HzRGs
is statistically significant.
This significance does not disappear by including QSO2s 
and Sy2s: the corresponding Spearman rank-order correlation
coefficient and its statistical significance become
$r_{\rm S} = -0.47$ and $p(r_{\rm S}) = 2.3 \times 10^{-4}$
($N_{\rm data} = 63$). This anti-correlation has also been 
shown by Allen et al. (1998) for a smaller HzRG sample.
We will discuss the interpretation of this trend in \S\S4.1.

\begin{figure}
\centering
\rotatebox{-90}{\includegraphics[width=5.5cm]{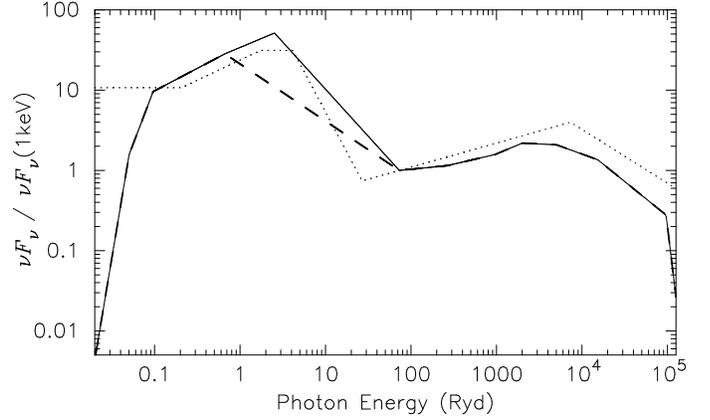}}
\caption{
   Adopted SEDs for our photoionization model calculations.
   Thin solid line denotes a SED with a strong UV bump and
   thick dashed line denotes a SED with a weak UV bump.
   Thin dotted line denotes the SED used by Mathews \& Ferland
   (1983). All the three SEDs are normalized to the flux at
   1 keV ($\simeq$73 Ryd).
}
\label{fig03}
\end{figure}

\begin{table}
\centering
\caption{Averaged Diagnostic Flux Ratios}
\label{table:3}
\begin{tabular}{l c c c}
\hline\hline
            \noalign{\smallskip}
  Sample                &  
  Number$^{\mathrm{a}}$ &
  C IV / He II &
  C III] / C IV \\
            \noalign{\smallskip}
\hline
            \noalign{\smallskip}
HzRG  & 49 & 1.42$^{+0.13}_{-0.08}$ & 0.67$^{+0.08}_{-0.10}$ \\
            \noalign{\smallskip}
QSO2  &  6 & 2.04$^{+0.42}_{-0.28}$ & 0.57$^{+0.05}_{-0.04}$ \\   
            \noalign{\smallskip}
Sy2   &  8 & 2.20$^{+0.17}_{-0.14}$ & 0.49$^{+0.06}_{-0.05}$ \\   
            \noalign{\smallskip}
\hline
\end{tabular}
\begin{list}{}{}
\item[$^{\mathrm{a}}$]  
  Number of objects for which both of the flux ratios
  of C{\sc iv}$\lambda$1549/He{\sc ii}$\lambda$1640
  and C{\sc iii}]$\lambda$1909/C{\sc iv}$\lambda$1549
  were measured. Objects with a lower-limit or
  upper-limit data are not included.
\end{list}
\end{table}

\section{Photoionization models}

To provide a quantitative interpretation of Figure 2,
we carried out
photoionization model calculations. The method and
the results are given below.

\subsection{Model method}

\begin{figure*}
\centering
\rotatebox{-90}{\includegraphics[width=11.3cm]{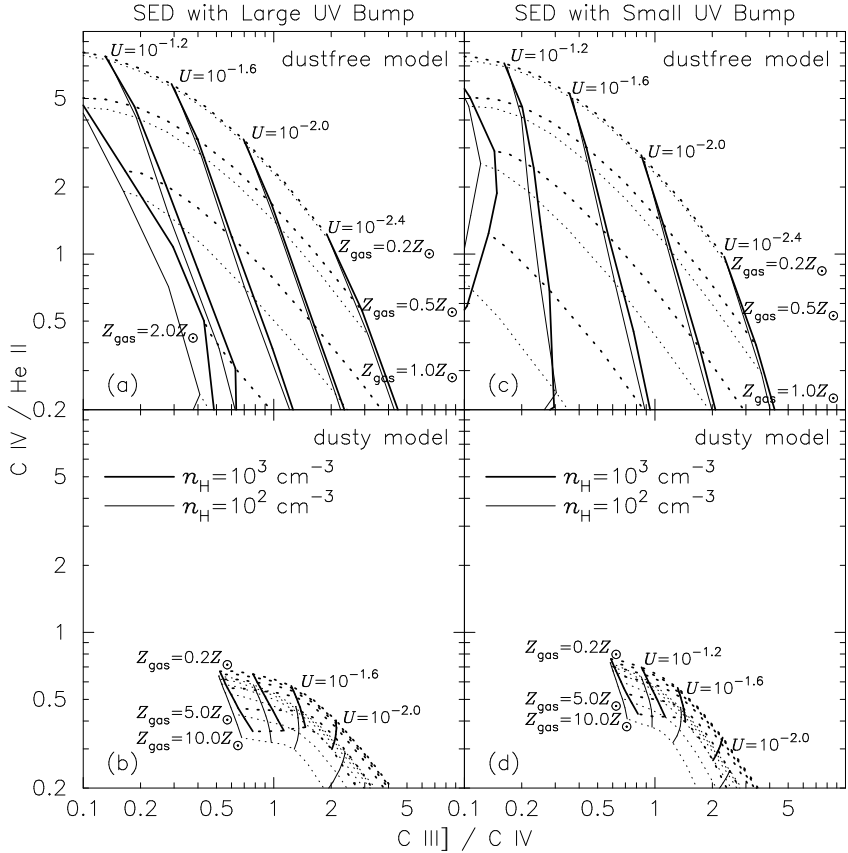}}
\caption{
   Model results plotted on a diagram of 
   C{\sc iv}$\lambda$1549/He{\sc ii}$\lambda$1640 versus
   C{\sc iii}]$\lambda$1909/C{\sc iv}$\lambda$1549.
   Models for gas clouds with $n_{\rm H} = 10^2$ cm$^{-3}$
   (thin lines) and $10^3$ cm$^{-3}$ (thick lines) are presented.
   Constant-$Z_{\rm gas}$ sequences and constant-$U$
   sequences are denoted by dotted and solid lines, respectively.
   (a) Dust-free models adopting the input SED with a large UV bump.
   (b) Dusty models adopting the input SED with a large UV bump.
   (c) Dust-free models adopting the input SED with a small UV bump.
   (d) Dusty models adopting the input SED with a small UV bump.
}
\label{fig04}
\end{figure*}

We performed model runs by using the public photoionization 
code Cloudy version 94\footnote{
We confirmed for some models that the results of the 
calculations do not change significantly if using
Cloudy version 96 instead of version 94; the difference
is $\sim 10$\% at maximum.} 
(Ferland 1997; Ferland et al. 1998). 
For simplicity, we assumed uniform gas density clouds with a 
plane-parallel geometry, and we examined both 
dust-free and dusty cases. For the models with dust, 
Orion-type graphite and silicate grains (Baldwin et al. 1991; 
Ferland 1997) were included. Note that the calculations are
not sensitive to the assumption on the grain composition
(\S\S4.1.2).
Effects of dust grains on the depletion of heavy elements
and on the radiative transfer were consistently treated by Cloudy.
The parameters for the calculations are (1) the spectral energy 
distribution (SED) of the photoionizing continuum radiation; 
(2) the hydrogen density of a cloud ($n_{\rm H}$); 
(3) the ionization parameter ($U$), i.e., the ratio of the 
ionizing photon density to the hydrogen density at the 
irradiated surface of a cloud; 
(4) the column density of a cloud ($N_{\rm H}$); and 
(5) the elemental composition of the gas.

As for the SED of the ionizing photons, two extreme cases of 
SED were investigated. The first one is a SED with a weak UV
thermal bump, which matches the HST QSO templates
(Zheng et al. 1998; Telfer et al. 2002; see Marconi et al. 2004 
for more details). The second one has a strong UV thermal bump
to match the QSO template by Scott et al. (2004). Both 
SEDs have the same optical to X-ray ratio $\alpha_{\rm OX}$ 
(Zamorani et al. 1981),
i.e., $\alpha_{\rm OX} = -1.49$ (see Elvis et al. 2002), 
but different slopes in
the energy range of $9.1{\rm eV} \leq h\nu \leq 35.5{\rm eV}$;
$\alpha = -2.17$ for the SED with a strong UV bump and
$\alpha = -1.71$ for the SED with a weaker UV bump, where
$f_\nu \propto \nu^\alpha$.
See Figure 3 for a graphical representation of the two SEDs.
Just for the readers' convenience, in Figure 3 we also show 
the SED deduced by Mathews \& Ferland 
(1983) that has been sometimes used for photoionization model 
calculations (see Ferland 1997), although we do not use this 
SED in this work.

\begin{figure*}
\centering
\rotatebox{-90}{\includegraphics[width=11.3cm]{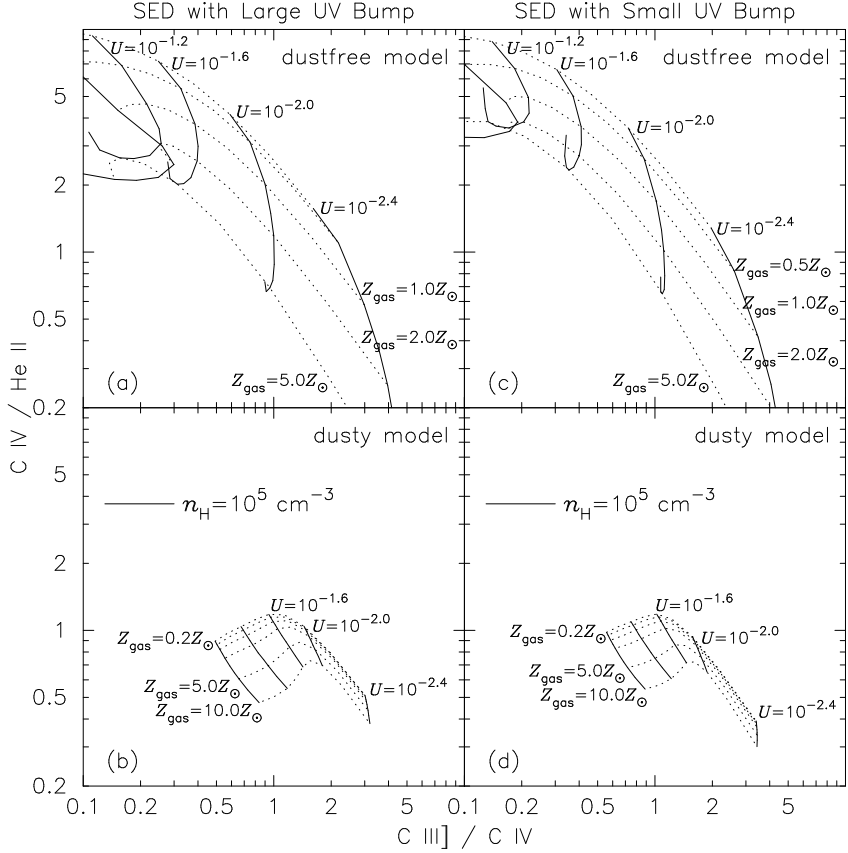}}
\caption{
   Same as Figure 4 but models with $n_{\rm H} = 10^5$ cm$^{-3}$
   are plotted. In panels (a) and (c), constant-$Z_{\rm gas}$ 
   sequences (dotted lines) are plotted for $Z_{\rm gas} =$
   0.2, 0.5, 1.0, 2.0 and 5.0 $Z_\odot$ (but not for 
   $Z_{\rm gas} = 10.0 Z_\odot$), although constant-$U$
   sequences (solid lines) are plotted for the range
   $0.2 \leq Z_{\rm gas}/Z_\odot < 10.0$.
}
\label{fig05}
\end{figure*}

We investigated gas clouds with gas densities 
$n_{\rm H} = 10^2$ cm$^{-3}$, $10^3$ cm$^{-3}$,
$10^5$ cm$^{-3}$ and $10^6$ cm$^{-3}$, and
ionization parameters $U = 10^{-2.8} - 10^{-0.8}$, 
as presented in \S\S3.2. The column 
density $N_{\rm H}$ was determined by the criterion that the
calculations for dust-free gas clouds were stopped when the 
temperature falls to 100K, below which the gas does not 
contribute significantly to the observed optical emission 
line spectra. Although this lower-limit temperature is much 
lower than other calculations for ionization-bounded clouds 
in the literature, this criterion is necessary to calculate
low-density dust-free gas clouds with a high metallicity 
because the equilibrium temperature of such gas clouds is 
sometimes lower than 1000K, as it will be shown later (see 
also Ferland et al. 1984). For models with dust grains, the 
stopping temperature was set to 2000K. This is because the
gas temperature does not decrease efficiently in 
partially-ionized regions when clouds contain dust, due 
mainly to the depletion of heavy elements (i.e., coolants) 
and to the heating by photoelectrons emitted from grains 
(see, e.g., 
Shields \& Kennicutt 1995 for details). Note that the 
results of our calculations are not sensitive to the 
lower-limit temperature because we are concerned only in
relatively high-ionization emission lines, which arise in 
fully-ionized regions, and not within partially-ionized 
regions. For the chemical composition of dust-free gas 
clouds, we assumed that the all metals scale by keeping 
solar proportions except for nitrogen, that scales as the 
square power of other metal abundances, because nitrogen 
is a secondary element (see, e.g., Hamann et al. 2002). 
Here the solar elemental abundances are taken from 
Grevesse \& Anders (1989) with extensions by 
Grevesse \& Noels (1993). For dusty gas clouds, we assumed 
the depleted gas-phase elemental abundance ratios by 
adopting the depletion factors given by Ferland (1997).

\begin{figure*}
\centering
\includegraphics[width=11.0cm]{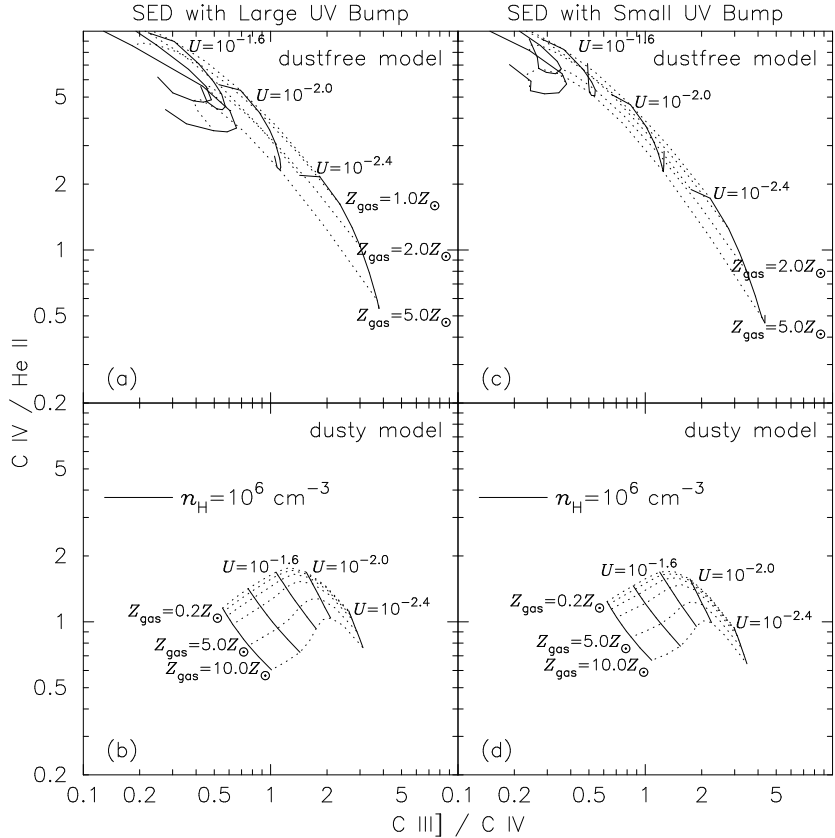}
\caption{
   Same as Figure 5 but models with $n_{\rm H} = 10^6$ cm$^{-3}$
   are plotted. 
}
\label{fig06}
\end{figure*}

\subsection{Model results}

In Figures 4, 5 and 6, the results of the model 
calculations are plotted on a diagram with 
C{\sc iv}$\lambda$1549/He{\sc ii}$\lambda$1640 and 
C{\sc iii}]$\lambda$1909/C{\sc iv}$\lambda$1549,
for both dust-free and dusty cases. Models for low-density 
gas clouds ($n_{\rm H} = 10^2$ cm$^{-3}$ and 
$10^3$ cm$^{-3}$) are presented in Figure 4, those for 
high-density gas clouds ($n_{\rm H} = 10^5$ cm$^{-3}$)
are presented in Figure 5, and those for very high-density
gas clouds ($n_{\rm H} = 10^6$ cm$^{-3}$) are presented
in Figure 6. For low-density cases, the 
difference of the model results between
those with $n_{\rm H} = 10^2$ cm$^{-3}$ and those with
$n_{\rm H} = 10^3$ cm$^{-3}$ is very small.

For low-density gas clouds without dust grains, the flux 
ratio of C{\sc iv}$\lambda$1549/He{\sc ii}$\lambda$1640 
strongly depends on the gas metallicity while 
C{\sc iii}]$\lambda$1909/C{\sc iv}$\lambda$1549 allows to
remove the degeneracy with $U$, and thus the diagram
with C{\sc iv}$\lambda$1549/He{\sc ii}$\lambda$1640 and 
C{\sc iii}]$\lambda$1909/C{\sc iv}$\lambda$1549 is a good 
metallicity diagnostics (Figures 4a and 4c). 
The strong variation of
C{\sc iv}$\lambda$1549/He{\sc ii}$\lambda$1640 with 
metallicity is due to the strong dependence of 
C{\sc iv}$\lambda$1549 emissivity on the gas temperature, 
which decreases rapidly with metallicity (in dust-free
clouds) due to an efficient cooling by infrared 
fine-structure lines (e.g., Ferland et al. 1984).
The metallicity dependences of
some infrared fine-structure lines ([O{\sc iii}]88$\mu$m, 
[N{\sc iii}]57$\mu$m and [Ne{\sc iii}]15.6$\mu$m) and
gas temperature for gas clouds with $n_{\rm H} = 10^2$
cm$^{-3}$ are shown in Figure 7. Here we adopt the averaged
temperature over doubly-ionized helium regions as a
representative temperature in photoionization equilibrium.
The fluxes of the fine-structure lines plotted in Figure 7
are normalized to the H$\beta$ flux, because the H$\beta$
flux scales with the ionization photon flux.

The metallicity dependence of 
C{\sc iv}$\lambda$1549/He{\sc ii}$\lambda$1640 in
dust-free gas clouds becomes smaller at
higher gas densities (Figures 5a, 5c, 6a, and 6c).
This is mainly due to the decreased cooling efficiency of
infrared fine-structure lines, which are suppressed in
high-density gas clouds owing to collisional de-excitation
(Figure 7). However, even when gas clouds with
$n_{\rm H} = 10^5$ cm$^{-3}$ are concerned, the diagram 
with C{\sc iv}$\lambda$1549/He{\sc ii}$\lambda$1640 and 
C{\sc iii}]$\lambda$1909/C{\sc iv}$\lambda$1549 maybe
useful to constrain the gas metallicity, because 
the C{\sc iv}$\lambda$1549/He{\sc ii}$\lambda$1640 ratio
depends on metallicity even at $n_{\rm H} = 10^5$ cm$^{-3}$, 
although the inferred metallicity is different depending on
the specific gas density (Figures 5a and 5c).
The C{\sc iv}$\lambda$1549/He{\sc ii}$\lambda$1640 versus
C{\sc iii}]$\lambda$1909/C{\sc iv}$\lambda$1549 diagram is
no more useful when the gas density is very high,
$n_{\rm H} = 10^6$ cm$^{-3}$ (Figures 6a and 6c). The
metallicity dependence of the
C{\sc iv}$\lambda$1549/He{\sc ii}$\lambda$1640 has almost
disappeared at this gas density. However, as shown later,
the very high-density models ($n_{\rm H} = 10^6$ cm$^{-3}$)
do not provide a good description of the observed data. 

Gas clouds with dust grains show only a very small dependence 
of C{\sc iv}$\lambda$1549/He{\sc ii}$\lambda$1640 on 
metallicity (Figures 4b, 4d, 5b, 5d, 6b, and 6d).
This is mainly because the equilibrium temperature of gas 
clouds does not drop off significantly when gas metallicity 
is high, thanks to the presence of dust grains (Figure 7d;
see Shields \& Kennicutt 1995 for more details). This 
result is almost independent of the adopted SED, 
gas density, and ionization parameter.

Figures 4, 5 and 6 indicate that the results 
with a large UV bump SED and 
with a small UV bump SED are similar. We thus conclude that
SED effects on our analysis and discussion are negligible.
In the following discussion, we only refer to the models
with a small UV bump SED. 

Note that the diagnostic diagram on which we are focusing has 
been investigated for various purposes in the past, since
all the three lines (He{\sc ii}$\lambda$1640, 
C{\sc iv}$\lambda$1549 and C{\sc iii}]$\lambda$1909) are
very strong and the wavelength separation is small (i.e.,
their ratios are little sensitive to dust extinction). 
Allen et al. (1998)
investigated photoionization models and fast-shock models
showing that the data of HzRGs on the
C{\sc iv}$\lambda$1549/He{\sc ii}$\lambda$1640 versus
C{\sc iii}]$\lambda$1909/C{\sc iv}$\lambda$1549 diagram are
consistent with photoionization, but are hard to be accounted
for by fast-shock models. Groves et al. (2004) showed
that this diagram is useful to examine the gas metallicity
of NLRs and mentioned that HzRGs may have gas clouds with
sub-solar metallicity. We use this diagram to analyze the
gas metallicity of NLRs for a larger sample of high-$z$
narrow-line AGNs and to investigate the possible metallicity
evolution of NLRs.

Finally we tackle the issue of whether the flux of 
C{\sc iii}]$\lambda$1909 is contaminated by the 
Si{\sc iii}]$\lambda\lambda$1883,1892 emission significantly.
The predicted ratio of
Si{\sc iii}]$\lambda\lambda$1883,1892/C{\sc iii}]$\lambda$1909
is plotted as a function of ionization parameter in Figure 8.
Here we investigate gas clouds with ($n_{\rm H}$, $Z_{\rm gas}$) 
= ($10^2$ cm$^{-3}$, 0.5 $Z_\odot$),
($10^2$ cm$^{-3}$, 2.0 $Z_\odot$),
($10^5$ cm$^{-3}$, 0.5 $Z_\odot$), and
($10^5$ cm$^{-3}$, 2.0 $Z_\odot$).
For clouds with dust grains, the predicted ratio is $\sim$0.01
regardless of density, metallicity, and ionization parameter.
The contribution of Si{\sc iii}]$\lambda\lambda$1883,1892 is
thus negligible when dusty clouds are concerned.
For dust-free clouds, the predicted ratios are higher,
i.e., $\sim$0.1. However, this is still 
significantly smaller than the measurement errors on
C{\sc iii}]$\lambda$1909 fluxes.
Therefore, we conclude that the contribution of
Si{\sc iii}]$\lambda\lambda$1883,1892 does not affect our
results and discussion significantly.

\begin{figure*}
\centering
\includegraphics[width=11.0cm]{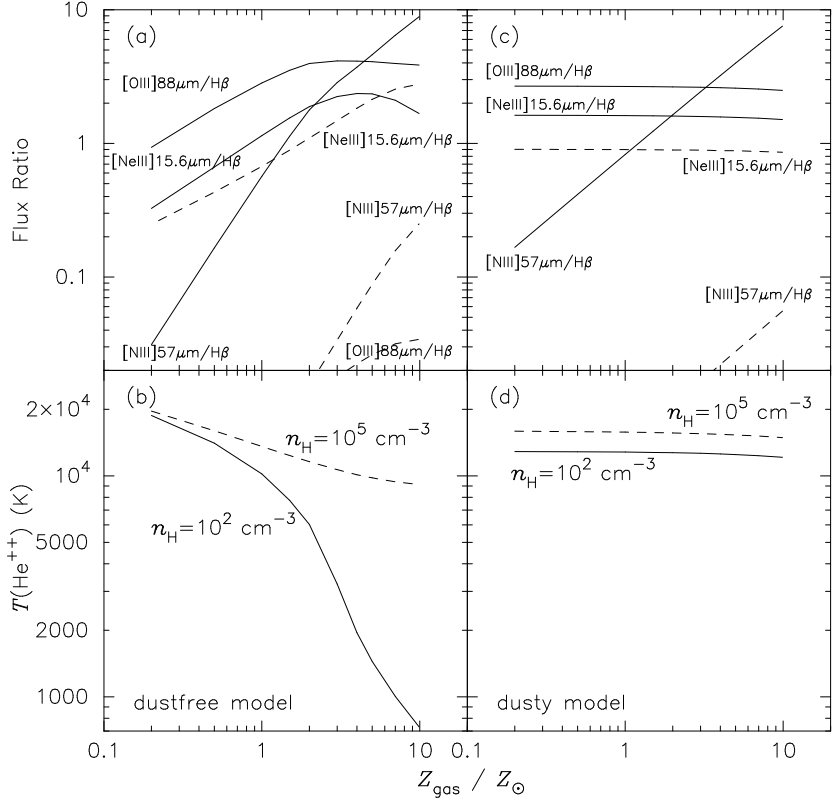}
\caption{
(a) Predicted flux ratios of [O{\sc iii}]88$\mu$m/H$\beta$,
    [N{\sc iii}]57$\mu$m/H$\beta$ and 
    [Ne{\sc iii}]15.6$\mu$m/H$\beta$ as functions of metallicity, 
    from dust-free gas clouds 
    with $n_{\rm H} = 10^2$ cm$^{-3}$ (solid lines) and those 
    with $n_{\rm H} = 10^5$ cm$^{-3}$ (dashed lines). The 
    ionization parameter of $U = 10^{-2.0}$ and the input SED 
    with a small UV bump are assumed.
(b) Averaged temperature of doubly-ionized helium regions of 
    dust-free gas clouds with $n_{\rm H} = 10^2$ cm$^{-3}$
    (solid line) and those with $n_{\rm H} = 10^5$ cm$^{-3}$
    (dashed line) as a function of metallicity. 
    The ionization parameter and the adopted SED
    are the same as (a).
(c) Same as (a) but for clouds with dust grains. The predicted
    flux ratio of [O{\sc iii}]88$\mu$m/H$\beta$ is too low
    ($<$ 0.01) and thus not presented in the panel.
(d) Same as (b) but for clouds with dust grains.
}
\label{fig07}
\end{figure*}

\begin{figure}
\centering
\rotatebox{-90}{\includegraphics[width=10.5cm]{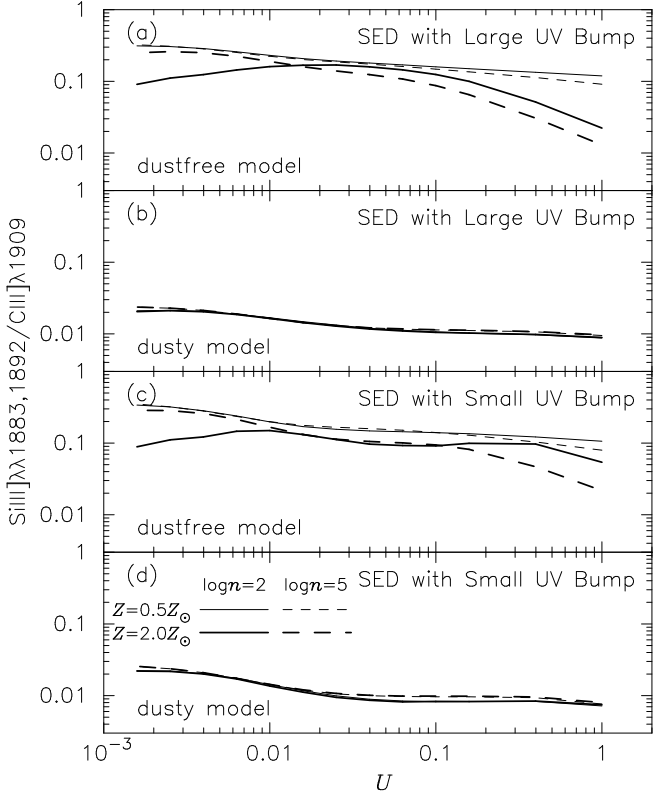}}
\caption{
   Predicted flux ratio of 
   Si{\sc iii}]$\lambda\lambda$1883,1892/C{\sc iii}]$\lambda$1909
   as a function of ionization parameter. Solid and dashed lines
   denote the models with $n_{\rm H} = 10^2$ cm$^{-3}$ and
   $n_{\rm H} = 10^5$ cm$^{-3}$, respectively.
   Thin and thick lines denote the models with 
   $Z_{\rm gas} = 0.5 Z_\odot$ and $Z_{\rm gas} = 2.0 Z_\odot$,
   respectively.
   (a) Dust-free models with a large UV bump SED.
   (b) Dusty models with a large UV bump SED.
   (c) Dust-free models with a small UV bump SED.
   (d) Dusty models with a small UV bump SED.
}
\label{fig08}
\end{figure}

\section{Discussion}

\subsection{Comparison of the data with models}

\subsubsection{Shock models}

\begin{figure}
\centering
\rotatebox{-90}{\includegraphics[width=8.5cm]{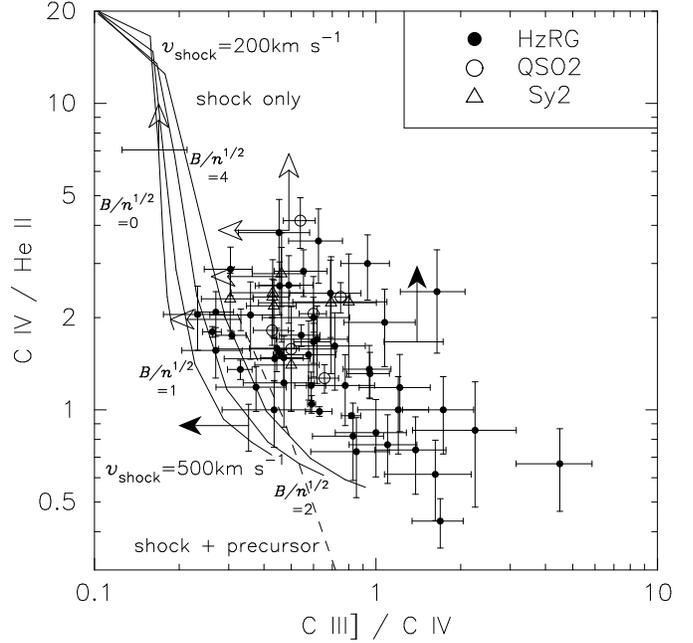}}
\caption{
   Same as Figure 2a but shock models of Dopita \& Sutherland (1996)
   are overplotted. Solid lines denote the predictions of
   pure shock-only models with magnetic parameters of
   0, 1, 2, and 4$\mu$G cm$^{-3/2}$ (from left to right in the panel).
   Dashed line denotes the predictions of shock plus precursor models.
}
\label{fig09}
\end{figure}

Before comparing the data presented in \S2 with
the photoionization models presented in \S3, we examine
whether shock models can explain the observed flux ratios.
In Figure 9, we compare the
data of the HzRG, QSO2 and Sy2 samples with the shock models
presented by Dopita \& Sutherland (1996).
Both pure shock-only models and shock plus precursor models
are examined here: shock-only models with a shock velocity of
150 km s$^{-1}$ $\leq$ $v_{\rm shock}$ $\leq$ 500 km s$^{-1}$
and with a magnetic parameter of 0 $\mu$G cm$^{-3/2}$ $\leq$ 
$B/n^{1/2}$ $\leq$ 4 $\mu$G cm$^{-3/2}$,
and shock plus precursor models with a shock velocity of
200 km s$^{-1}$ $\leq$ $v_{\rm shock}$ $\leq$ 500 km s$^{-1}$.
As shown in Figure 9, both pure shock-only models and
shock plus precursor models underpredict the flux ratio of
C{\sc iii}]$\lambda$1909/C{\sc iv}$\lambda$1549 and thus
cannot explain the observed data. This suggests that
the main ionization mechanism of NLRs (or at least the
C{\sc iv}$\lambda$1549, He{\sc ii}$\lambda$1640 and 
C{\sc iii}]$\lambda$1909 emitting regions in NLRs) is not 
associated with shocks, but is photoionization. Figure 9
also suggests that the difference in the flux ratios 
between HzRGs and QSO2s -- Sy2s cannot be ascribed to 
shocks. Instead, a preferential contribution of shocks to QSO2s 
(not to HzRGs) is required if shocks are at the origin of the 
difference in the flux ratios. This requirement is in the 
opposite direction of the natural expectation (HzRGs should
be more affected by shocks because of jets and expanding
radio lobes). Therefore, the systematic difference in the 
flux ratios between HzRGs and QSO2s -- Sy2s must be ascribed
to causes other than shocks. This issue will be discussed
further in \S\S4.1.3.

\subsubsection{Dust grains}

In Figure 10, we compare the observational data with the 
results of our photoionization model calculations, both
with and without dust grains. Contrary to the dust-free 
models, the dusty models predict too narrow ranges of the
C{\sc iv}$\lambda$1549/He{\sc ii}$\lambda$1640 flux ratio,
regardless of the gas density (Figures 10b, 10d, and 10f).
The C{\sc iv}$\lambda$1549/He{\sc ii}$\lambda$1640 flux 
ratio varies only in a factor of 3 at maximum even when
the metallicity varies in the range 
$0.2 \leq Z_{\rm gas}/Z_\odot \leq 10.0$.
More importantly, the dusty models 
cannot explain values of
C{\sc iv}$\lambda$1549/He{\sc ii}$\lambda$1640 larger 
than $\sim 1.5$, which are instead observed in 
most sources. These results suggest that the 
C{\sc iv}$\lambda$1549, 
He{\sc ii}$\lambda$1640 and C{\sc iii}]$\lambda$1909 
emitting regions in NLRs are dust-free. This is consistent 
with previous works that gas clouds in the high-ionization 
part of NLRs are dust-free (e.g., Marconi et al. 1994;
Ferguson et al. 1997; Nagao et al. 2003).
We verified that the effects of changing the grain composition
(see Ferland 1997 for details) are less than 30\% on the line flux ratios. 
We thus conclude that the models with dust
grains are not appropriate to interpret the ratios of
C{\sc iv}$\lambda$1549/He{\sc ii}$\lambda$1640 and 
C{\sc iii}]$\lambda$1909/C{\sc iv}$\lambda$1549.

\subsubsection{Ionization parameter}

In the case of dust-free models, the models with low
density ($n_{\rm H} = 10^2$ cm$^{-3}$ and $10^3$ cm$^{-3}$) 
and high density ($n_{\rm H} = 10^5$ cm$^{-3}$) can 
successfully explain the observed range of flux ratios
C{\sc iv}$\lambda$1549/He{\sc ii}$\lambda$1640 and 
C{\sc iii}]$\lambda$1909/C{\sc iv}$\lambda$1549
(Figures 10a and 10c), while the very high-density 
models ($n_{\rm H} = 10^6$ cm$^{-3}$) cannot explain most
of the observational data (Figure 10e).
And we can deduce that $n_{\rm H} = 10^6$ cm$^{-3}$ is an
upper bound to the average NLR density.
Both high-density models and low-density models suggest
that the observational data are consistent with clouds
with an ionization parameter 
$10^{-2.2} \la U \la 10^{-1.4}$ for HzRGs and
$U \approx 10^{-1.6}$ for QSO2s and Sy2s (Figures 10a and 
10c). The model sequences with a constant ionization 
parameter have negative slope in the diagram of 
C{\sc iv}$\lambda$1549/He{\sc ii}$\lambda$1640 versus
C{\sc iii}]$\lambda$1909/C{\sc iv}$\lambda$1549, which 
is consistent with the observational trend described in
\S\S2.2. Therefore the anti-correlation 
between C{\sc iv}$\lambda$1549/He{\sc ii}$\lambda$1640 and
C{\sc iii}]$\lambda$1909/C{\sc iv}$\lambda$1549 seen in HzRGs
can be naturally explained by the wide range of the ionization
parameter for HzRGs. Although the inferred range of ionization
parameter is higher than the range estimated through
rest-frame optical diagnostics (which use lower ionization 
emission lines) reported by some previous works (e.g., 
Ferland \& Netzer 1983; Ho et al. 1993), it is consistent with
previous estimates of the ionization parameter for the
high-ionization parts in NLRs (e.g., Oliva et al. 1994;
Murayama \& Taniguchi 1998; Nagao et al. 2001a, 2001b).

\subsubsection{Gas density and metallicity}

\begin{figure*}
\centering
\rotatebox{-90}{\includegraphics[width=10.8cm]{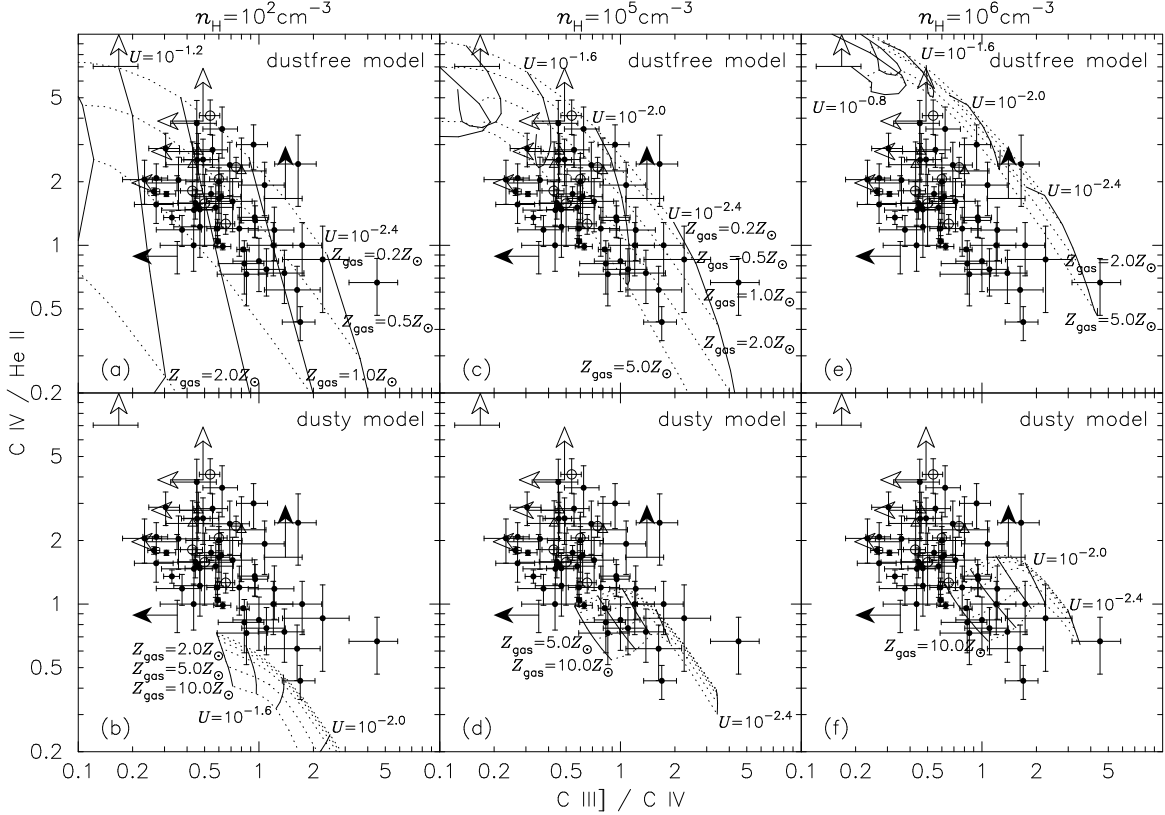}}
\caption{
   Observational data and model results are plotted on a diagram 
   of C{\sc iv}$\lambda$1549/He{\sc ii}$\lambda$1640 versus
   C{\sc iii}]$\lambda$1909/C{\sc iv}$\lambda$1549. A SED with a
   small UV bump is adopted in the models. Symbols are the same as
   those in Figure 2.
   Dust-free models are plotted in the upper panels (a, c, e) while
   dusty models are plotted in the lower panels (b, d, f).
   Adopted gas densities are $n_{\rm H} = 10^2$ cm$^{-3}$ (panels a, b),
   $n_{\rm H} = 10^5$ cm$^{-3}$ (panels c, d), and
   $n_{\rm H} = 10^6$ cm$^{-3}$ (panels e, f).
}
\label{fig10}
\end{figure*}

As mentioned in the last subsection, photoionization models
with $n_{\rm H} = 10^6$ cm$^{-3}$ predict too narrow
ranges of the flux ratios of 
C{\sc iv}$\lambda$1549/He{\sc ii}$\lambda$1640 and 
C{\sc iii}]$\lambda$1909/C{\sc iv}$\lambda$1549 with respect
to the observed ranges. This suggests that the typical density
of the C{\sc iv}$\lambda$1549, He{\sc ii}$\lambda$1640 and
C{\sc iii}]$\lambda$1909-emitting region in the NLR should
be less than $10^6$ cm$^{-3}$. Therefore, in the following 
discussion, we consider only the two models, i.e.,
the low-density models with $n_{\rm H}= 10^2$ cm$^{-3}$ (note
that the results are very similar if adopting 
$n_{\rm H}= 10^3$ cm$^{-3}$) and the high-density models
with $n_{\rm H} = 10^5$ cm$^{-3}$.

As shown in Figures 10a and 10c, the estimated metallicity 
is different when different gas densities are adopted. 
Models with $n_{\rm H} = 10^2$ cm$^{-3}$ suggest sub-solar 
metallicities 
($0.2 \la Z_{\rm gas}/Z_\odot \la 1.0$; Figure 10a) 
while models with $n_{\rm H} = 10^5$ cm$^{-3}$ 
suggest a wide metallicity range 
($0.2 \la Z_{\rm gas}/Z_\odot \la 5.0$; Figure 10c).
The ``low-density and sub-solar metallicity'' scenario
appears to be consistent with the results reported by
Iwamuro et al. (2003), while the ``high-density with wide
metallicity distribution'' scenario appears in agreement
with the results reported by De Breuck et al. (2000).
Although some observational data with lower 
C{\sc iv}$\lambda$1549/He{\sc ii}$\lambda$1640 and 
C{\sc iii}]$\lambda$1909/C{\sc iv}$\lambda$1549 ratios
appear to deviate from the model predictions with
$n_{\rm H} = 10^5$ cm$^{-3}$ (Figure 10c), these deviations
can be reconciled by introducing a moderate amount of 
extinction. 
Although the above two scenarios are hard to be 
discriminated only by means of the 
C{\sc iv}$\lambda$1549/He{\sc ii}$\lambda$1640 versus 
C{\sc iii}]$\lambda$1909/C{\sc iv}$\lambda$1549 diagram, 
the actual situation may be intermediate
between the two possible scenarios (\S\S4.2).
Note that the absolute values of the inferred 
gas metallicity are not very accurate due to a density 
dependence of the flux ratio
C{\sc iv}$\lambda$1549/He{\sc ii}$\lambda$1640 at
high-densities ($n_{\rm H} \ga 10^4$ cm$^{-3}$). 
Nevertheless the diagnostic diagram in Figure 10 is useful if 
we are interested in the relative metallicity trends of NLR 
clouds, or in the evolution of gas metallicity, adopting the 
assumption that there are no strong correlations between gas 
density and redshift.
Finally, we note that the density affects the inferred 
metallicity mostly at $Z_{\rm gas} \ga 0.5 Z_\odot$.
At low metallicities ($Z_{\rm gas} \sim 0.2 Z_\odot$)
the models are less sensitive to the gas density 
(Figures 10a and 10c).
This result is important, since it allows to use this
diagnostic diagram to identify low metallicity NLRs, 
regardless of the gas density.

\begin{figure*}
\centering
\includegraphics[width=11.0cm]{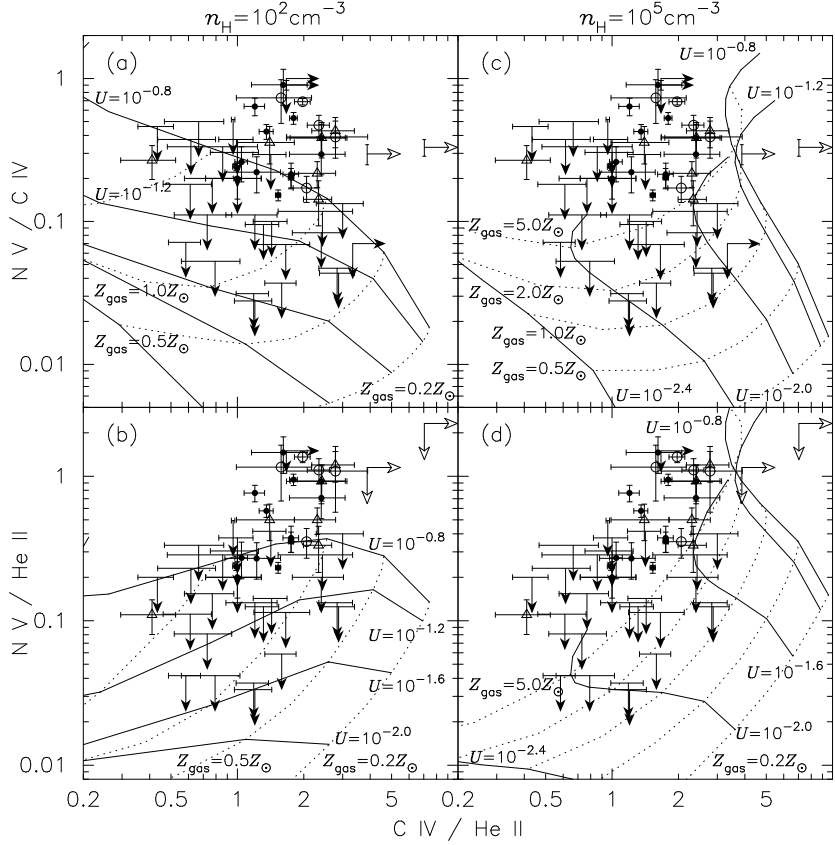}
\caption{
   Observed and predicted flux ratios. Dust-free models with a small
   UV bump are plotted. Symbols are the same as those in Figure 2.
   Observed data compared with models for $n_{\rm H} = 10^2$ cm$^{-3}$
   on a diagram of N{\sc v}$\lambda$1240/C{\sc iv}$\lambda$1549 
   versus C{\sc iv}$\lambda$1549/He{\sc ii}$\lambda$1640 (a) and on
   a diagram of N{\sc v}$\lambda$1240/He{\sc ii}$\lambda$1640 versus
   C{\sc iv}$\lambda$1549/He{\sc ii}$\lambda$1640 (b). Data compared
   with models for $n_{\rm H} = 10^5$ cm$^{-3}$ on
   N{\sc v}$\lambda$1240/C{\sc iv}$\lambda$1549 
   versus C{\sc iv}$\lambda$1549/He{\sc ii}$\lambda$1640 (c) and on
   N{\sc v}$\lambda$1240/He{\sc ii}$\lambda$1640 versus
   C{\sc iv}$\lambda$1549/He{\sc ii}$\lambda$1640 (d).
}
\label{fig11}
\end{figure*}

\subsection{Comparison with the N{\sc v}$\lambda$1240 diagnostics}

The gas metallicity of the BLR clouds in QSOs has been 
often investigated by using the diagnostic flux ratios of
N{\sc v}$\lambda$1240/C{\sc iv}$\lambda$1549 and
N{\sc v}$\lambda$1240/He{\sc ii}$\lambda$1640 (e.g.,
Hamann \& Ferland 1992, 1999; Dietrich et al. 2003;
Nagao et al. 2005). This method has been sometimes applied
also to the NLR clouds (e.g., van Ojik et al. 1994;
Villar-Martin et al. 1999; De Breuck et al. 2000; 
Vernet et al. 2001). Therefore it is interesting to compare
our diagnostics with the N{\sc v}$\lambda$1240 diagnostics.
In Figure 11, the observational data are compared with the
photoionization models for $n_{\rm H} = 10^2$ cm$^{-3}$
and $n_{\rm H} = 10^5$ cm$^{-3}$ on the diagram of 
N{\sc v}$\lambda$1240/C{\sc iv}$\lambda$1549 versus
C{\sc iv}$\lambda$1549/He{\sc ii}$\lambda$1640, and
N{\sc v}$\lambda$1240/He{\sc ii}$\lambda$1640 versus
C{\sc iv}$\lambda$1549/He{\sc ii}$\lambda$1640.
Although high ratios of 
N{\sc v}$\lambda$1240/C{\sc iv}$\lambda$1549 and
N{\sc v}$\lambda$1240/He{\sc ii}$\lambda$1640 are sometimes
interpreted simply as a result of high metallicities,
the low-density models ($n_{\rm H} = 10^2$ cm$^{-3}$)
with a high metallicity predict not only high
N{\sc v}$\lambda$1240/C{\sc iv}$\lambda$1549 ratios but
also very low C{\sc iv}$\lambda$1549/He{\sc ii}$\lambda$1640
ratios, in disagreement with the observed data (Figure 11a).
This is mainly attributed to the decreased equilibrium
gas temperature due to the efficient cooling by infrared
fine-structure lines as investigated in \S\S3.2 (Figure 7).
Moreover the low-density models predict
N{\sc v}$\lambda$1240/He{\sc ii}$\lambda$1640 $<$ 0.4
when $U \leq 10^{-0.8}$ is concerned, which is also in 
disagreement with the observed data (Figure 11b).
The high-density models ($n_{\rm H} = 10^5$ cm$^{-3}$),
on the other hand, predict higher 
C{\sc iv}$\lambda$1549/He{\sc ii}$\lambda$1640 ratios than
the low-density models owing to the suppressed cooling.
Since the high-density models predict higher
C{\sc iv}$\lambda$1549/He{\sc ii}$\lambda$1640 ratios than
the observed values (Figures 11c and 11d), the typical
gas density of (the high-ionization part of) the NLRs
may be lower than $n_{\rm H} = 10^5$ cm$^{-3}$.

Although the N{\sc v}$\lambda$1240-detected objects appear 
to have super-solar metallicity, the N{\sc v}$\lambda$1240
fluxes of more than half of the objects in the sample are
upper-limits and they might be consistent with 
sub-solar metallicities. 
Summarizing, since N{\sc v}$\lambda$1240 becomes very
weak (generally undetected) for $Z_{\rm gas} \la 2 Z_\odot$
NLRs, it is very difficult to investigate the
metallicity evolution of the NLR based on
the N{\sc v}$\lambda$1240 diagnostics.

\begin{table*}
\centering
\caption{Averaged Diagnostic Flux Ratios of HzRGs}
\label{table:4}
\begin{tabular}{l c c c}
\hline\hline
            \noalign{\smallskip}
  Sample                &  
  Number$^{\mathrm{a}}$ &
  C IV / He II &
  C III] / C IV \\
            \noalign{\smallskip}
\hline
            \noalign{\smallskip}
HzRG $1.2 < z < 2.0$ 
  & 17 & 1.65$^{+0.24}_{-0.15}$ & 0.66$^{+0.10}_{-0.06}$ \\
HzRG $2.0 < z < 2.5$ 
  & 20 & 1.27$^{+0.15}_{-0.10}$ & 0.69$^{+0.14}_{-0.07}$ \\
HzRG $2.5 < z < 3.8$ 
  & 12 & 1.39$^{+0.31}_{-0.17}$ & 0.64$^{+0.22}_{-0.10}$ \\
HzRG 41.5 $<$ log $L$(He{\sc ii})$^{\mathrm{b}}$ $<$ 42.5
  & 13 & 1.72$^{+0.32}_{-0.19}$ & 0.89$^{+0.12}_{-0.08}$ \\
HzRG 42.5 $<$ log $L$(He{\sc ii})$^{\mathrm{b}}$ $<$ 43.0
  & 21 & 1.39$^{+0.18}_{-0.11}$ & 0.62$^{+0.15}_{-0.07}$ \\
HzRG 43.0 $<$ log $L$(He{\sc ii})$^{\mathrm{b}}$ $<$ 45.0
  & 15 & 1.26$^{+0.19}_{-0.12}$ & 0.58$^{+0.11}_{-0.06}$ \\
HzRG total 
  & 49 & 1.42$^{+0.13}_{-0.08}$ & 0.67$^{+0.08}_{-0.04}$ \\
            \noalign{\smallskip}
\hline
\end{tabular}
\begin{list}{}{}
\item[$^{\mathrm{a}}$]  
  Number of objects for which both of the flux ratios
  of C{\sc iv}$\lambda$1549/He{\sc ii}$\lambda$1640
  and C{\sc iii}]$\lambda$1909/C{\sc iv}$\lambda$1549
  were measured. Objects with a lower-limit or
  upper-limit data are not included.
\item[$^{\mathrm{b}}$]  
  Line luminosity of He{\sc ii}$\lambda$1640 in units 
  of ergs s$^{-1}$, calculated from the line flux
  given by De Breuck et al. (2000).
\end{list}
\end{table*}

\subsection{Metallicity evolution of the NLRs in AGNs}

Now we discuss whether the gas metallicity of the NLRs in 
high-$z$ AGNs evolves as a function of redshift. We first 
focus only on the HzRG data, because HzRGs and QSO2s may 
have a different NLR properties, as discussed above. 
To investigate this issue, we divide the sample of 
objects having a measurement of both
C{\sc iv}$\lambda$1549/He{\sc ii}$\lambda$1640 and
C{\sc iii}]$\lambda$1909/C{\sc iv}$\lambda$1549 (49 
objects, discarding the objects with upper-limits or
lower-limits) into three groups: $1.2 < z < 2.0$ (17 
objects), $2.0 < z < 2.5$ (20 objects), and $2.5 < z < 3.8$
(12 objects). Note that most of the highest-$z$ sample are 
at $2.5 < z < 3.0$ and only two objects are at $z > 3$
(Figure 1a). We calculate the logarithmically 
averaged flux ratios of 
C{\sc iv}$\lambda$1549/He{\sc ii}$\lambda$1640 and
C{\sc iii}]$\lambda$1909/C{\sc iv}$\lambda$1549 for 
these three groups. The results are given in Table 4 
and plotted on the diagnostic diagram in Figures 12a 
and 12c. The thick errorbars give the estimated errors on
the means. As clearly seen, no evolutionary tendency in 
the flux ratios within our HzRG sample is found. 
The higher-$z$ objects do not show 
significant metallicity decrease with respect to the 
lower-$z$ objects, at variance with the result reported
by De Breuck et al. (2000). Although the absolute value 
of the gas metallicity derived from Figure 10 is subject 
to non-negligible uncertainties, as discussed in 
\S\S4.1.4, our conclusion is not affected by this issue
in terms of relative gas metallicities, 
i.e., the gas metallicity in NLRs of HzRGs does not 
change significantly in the redshift range $1.2 < z < 3.8$,
or more conservatively, in the range $1.2 < z < 3$
(owing to the lack of objects at $z > 3$).

Recently Nagao et al. (2005) reported that the gas
metallicity of the BLR in QSOs with a given luminosity
is independent of redshift in the range $2.0 < z < 4.5$.
The latter result is consistent with that obtained for the
NLR metallicity of HzRGs presented in this
paper. Nagao et al. (2005) also reported that the BLR 
metallicity is tightly correlated with the QSO luminosity
(see also, e.g., Hamann \& Ferland 1993, 1999).
Motivated by this correlation between BLR metallicity
and QSO luminosity, we have investigated whether
the NLR metallicity is correlated with the AGN luminosity 
or not. However, it is very difficult to measure the
luminosity of both AGNs and their host galaxies when
type-2 AGNs are concerned. This is because the
central engine is hidden by the dusty torus and because 
the broad-band photometric flux is largely attributed to
the nebular emission, not only to the stellar continuum 
emission. We thus adopt the He{\sc ii}$\lambda$1640
emission-line luminosity [$L$(He{\sc ii})], as an
indicator of the AGN luminosity. This assumption
is based on the fact that the He{\sc ii}$\lambda$1640
luminosity is simply proportional to the volume of
the doubly-ionized helium region, which scales to the
AGN luminosity.
We calculate $L$(He{\sc ii}) from the 
He{\sc ii}$\lambda$1640 flux by
adopting a cosmology with ($\Omega_{\rm tot}$,
$\Omega_{\rm m}$, $\Omega_\Lambda$) = (1.0, 0.3, 0.7)
and $H_0$ = 70 km s$^{-1}$ cm$^{-1}$ Mpc$^{-1}$.
$L$(He{\sc ii}) is not corrected for the 
slit loss, which may be non-negligible for some cases.
To investigate the dependence of the line flux ratios on
$L$(He{\sc ii}), we divide HzRGs into three groups:
41.5 $<$ log $L$(He{\sc ii}) $<$ 42.5 (13 objects),
42.5 $<$ log $L$(He{\sc ii}) $<$ 43.0 (21 objects), and
43.0 $<$ log $L$(He{\sc ii}) $<$ 45.0 (15 objects), where
$L$(He{\sc ii}) is in units of erg s$^{-1}$.
The logarithmically averaged flux ratios and the RMS's
of C{\sc iv}$\lambda$1549/He{\sc ii}$\lambda$1640 
and C{\sc iii}]$\lambda$1909/C{\sc iv}$\lambda$1549 for 
these three groups are given in Table 4, and are plotted 
on the diagnostic diagram in Figure 12b and 12d.
The thick errorbars give the errors on the means.
As a result, we find a systematic trend in our HzRG sample
that the HzRGs with larger $L$(He{\sc ii}) (i.e.,
more luminous AGNs) tend to have lower
C{\sc iv}$\lambda$1549/He{\sc ii}$\lambda$1640 
and C{\sc iii}]$\lambda$1909/C{\sc iv}$\lambda$1549
flux ratios. This result is consistent to the interpretation
that the NLR in more luminous HzRGs have higher metallicity 
gas clouds. This ``luminosity-metallicity relation'' for
the NLR in HzRGs is in agreement with the same relation
seen for the BLRs in high-$z$ QSOs.

The similarity of the flux ratios between the 
QSO2 sample and the Sy2 sample may suggest that the gas 
metallicity of non-radio-selected AGNs does not evolve 
significantly from $z \sim 3$ to the local universe. 
However it should be kept in mind that there is a large 
difference in the absolute luminosity between the Sy2 
sample and the QSO2 sample. This result should be 
interpreted in the sense that the NLR metallicity is not
significantly different between faint local AGNs and
bright high-$z$ ($1.5 \la z \la 3.7$) AGNs.
If the luminosity-metallicity relation discussed above is
taken into account, this result might imply a metallicity
evolution of the NLR with a given luminosity from
high-$z$ to $z \sim 0$. Surveys and spectroscopic 
studies on type-2 faint AGNs at high-$z$ are required
to investigate this issue further.

\begin{figure*}
\centering
\includegraphics[width=11.0cm]{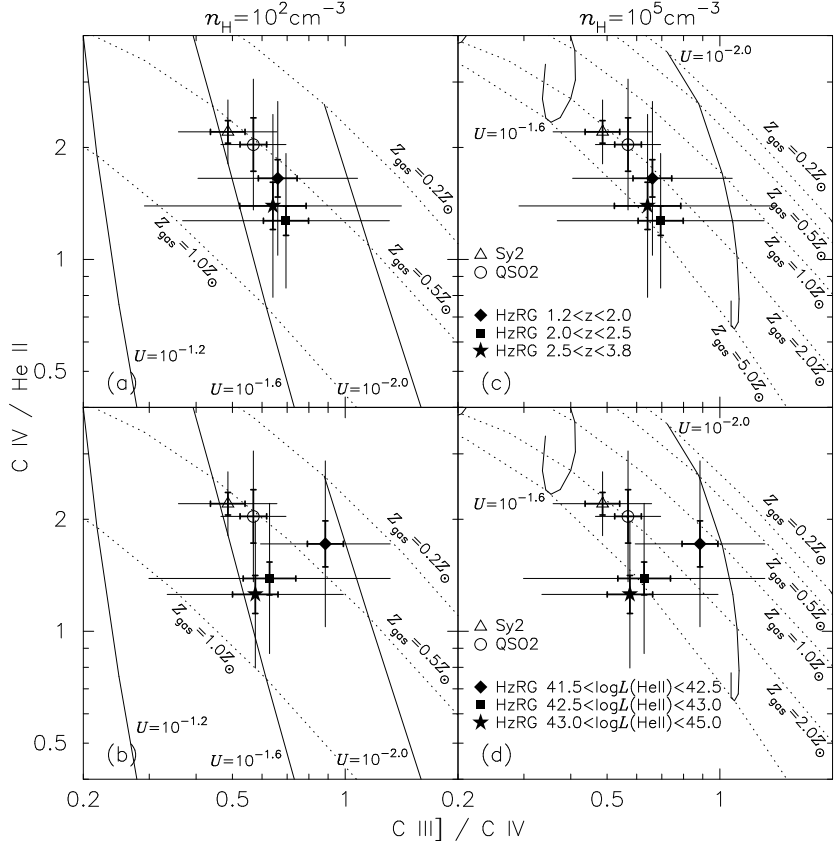}
\caption{
   Averaged flux ratios compared with the model predictions
   on the C{\sc iv}$\lambda$1549/He{\sc ii}$\lambda$1640 
   versus C{\sc iii}]$\lambda$1909/C{\sc iv}$\lambda$1549
   diagram. Filled circle, open circle and open triangle
   denote the averaged values of the HzRG sample, the
   QSO2 sample and Sy2s sample. Models with 
   $n_{\rm H} = 10^2$ cm$^{-3}$ are plotted in panels
   (a) and (b), while models with 
   $n_{\rm H} = 10^5$ cm$^{-3}$ are plotted in panels
   (c) and (d). For both cases, models with a small UV
   bump SED and without dust are plotted.
   The redshift dependence of the flux ratios of the HzRGs
   are investigated in panels (a) and (c), where filled
   diamond, square and star denote the averaged flux ratios 
   of the HzRGs at $1.2 < z < 2.0$, $2.0 < z < 2.5$ and
   $2.5 < z < 3.8$, respectively. The luminosity dependence 
   of the flux ratios of the HzRGs are investigated in panels 
   (b) and (d), where filled diamond, square and star denote 
   the averaged flux ratios of the HzRGs with
   41.5 $<$ log $L$(He{\sc ii}) $<$ 42.5, 
   42.5 $<$ log $L$(He{\sc ii}) $<$ 43.0, and
   43.0 $<$ log $L$(He{\sc ii}) $<$ 45.0 respectively, where
   $L$(He{\sc ii}) is in units of ergs s$^{-1}$.
   Error bars are as in Figure 2b.
}
\label{fig12}
\end{figure*}

It has been observationally confirmed that QSOs including 
ratio galaxies tend to reside in massive elliptical galaxies
at least in the local universe (e.g., McLure et al. 1999;
Dunlop et al. 2003; Floyd et al. 2004) and probably also 
in the high-$z$ universe (e.g., Falomo et al. 2005; 
Kuhlbrodt et al. 2005). Our results suggest that the main 
star-formation event in such massive host galaxies is 
completed. This picture is consistent with the
recent findings of massive evolved galaxies at $z > 1.5$
(e.g., Cimatti et al. 2004; McCarthy et al. 2004;
Labb\'{e} et al. 2005; Saracco et al. 2005).
The non-evolution of the gas metallicity of the NLR in our
sample suggests that the epoch of major chemical enrichment
in the host galaxies of QSO2s and HzRGs must have occurred
at $z \ga 3$. If the minimum timescale for 
significant enrichment of carbon ($\sim$0.5 Gyr) is
taken into account, the major epoch of star formation may be
at $z \ga 4$.

Finally we briefly discuss a specific, interesting QSO2, 
CDFS-901 (Figure 2a). Although the emission-line spectrum 
of this object is hard to be explained by photoionization 
models as seen in Figure 10, it appears to be consistent 
with pure shock-only models as shown in Figure 9. This 
may suggest that the NLR of CDFS-901 is dominated by 
shocks. However, there is another possibility, that is 
CDFS-901 may be a narrow-line type-1 QSO, i.e., a 
brighter analogue of narrow-line Seyfert 1 galaxies 
(NLS1s). If the emission-line spectrum of CDFS-901 is not 
from its NLR but from its BLR, its very large 
C{\sc iv}$\lambda$1549/He{\sc ii}$\lambda$1640 ratio 
($>7.0$) is naturally explained, since this flux ratio is 
expected to be $\sim 10$ for BLRs (although its 
C{\sc iii}]$\lambda$1909/C{\sc iv}$\lambda$1549 ratio 
$\sim$0.2 is very small for a BLR, e.g., Nagao et al. 2005).
Since the X-ray spectral slope (hardness ratio) of this 
object was not measured due to the lack of photon 
statistics, its nature remains ambiguous based on the 
currently available data. This object is interesting 
because NLS1s are sometimes thought to be AGNs with 
super-massive black holes in their growing-up phase
(e.g., Mathur 2000). If this object is a really high-$z$ 
analogue of NLS1s, it may be a very interesting target to
investigate the evolution of AGN activities and 
supermassive black holes.

\section{Summary}

In order to investigate the possible metallicity evolution 
of NLR gas clouds in AGNs, we compiled the fluxes of
C{\sc iv}$\lambda$1549, He{\sc ii}$\lambda$1640 and 
C{\sc iii}]$\lambda$1909 for a large sample of narrow-line
AGNs, including HzRGs, high-$z$ QSO2s, and local Sy2s. 
Since all of these three emission lines are moderately
strong even in the faint HzRGs and QSO2s, this approach
enables us to investigate a large number of such objects.
By comparing the compiled flux ratios with the results 
of our photoionization model calculations, we found the 
following results.
\begin{itemize}
  \item The observational data are inconsistent with the
        predictions of shock models, suggesting that
        the NLRs are mainly photoionized.
  \item The photoionization models with dust grains predict
        too narrow ranges of flux ratios and in disagreement 
        with the observed ranges, suggesting that the
        high-ionization part of NLRs (on which we focused in
        this work) is dust-free.
  \item The ionization parameter of (the high-ionization parts 
        of) NLRs is estimated to be 
        $10^{-2.2} \la U \la 10^{-1.4}$ for HzRGs 
        and $U \approx 10^{-1.6}$ for QSO2s and Sy2s. 
  \item The photoionization models with 
        $n_{\rm H} = 10^6$ cm$^{-3}$ cannot explain the
        observational data, suggesting that the typical gas
        density is lower than $10^6$ cm$^{-3}$.
  \item There are two possible interpretations for the observed
        data: low-density gas clouds 
        ($n_{\rm H} \la 10^3$ cm$^{-3}$) with a sub-solar 
        metallicity
        ($0.2 \la Z_{\rm gas}/Z_\odot \la 1.0$), or
        high-density gas clouds ($n_{\rm H} \sim 10^5$ cm$^{-3}$) 
        with a wide range of gas metallicity
        ($0.2 \la Z_{\rm gas}/Z_\odot \la 5.0$).
  \item Our method using only the flux ratios of
        C{\sc iv}$\lambda$1549/He{\sc ii}$\lambda$1640 and
        C{\sc iii}]$\lambda$1909/C{\sc iv}$\lambda$1549 is 
        particularly useful to examine relative difference in
        gas metallicity of NLR clouds, or to investigate
        possible metallicity evolution of NLRs, although
        the inferred absolute values of metallicity contain
        non-negligible uncertainties. 
  \item Regardless of the density, the proposed
        diagnostic diagram is useful to identify low metallicity
        NLRs ($Z_{\rm gas} \sim 0.2 Z_\odot$).
  \item We find no evidence suggesting a significant evolution 
        of the gas metallicity in the NLRs of HzRGs in the redshift
        range $1.2 \leq z \leq 3$.
  \item We find a systematic trend for more luminous AGNs to have more 
        metal-rich NLRs (luminosity-metallicity relation),
        which is in agreement with the results from the
        studies on the BLRs.
  \item The non-evolution of the gas metallicity of the NLRs
        implies that the major epoch of star formation in
        the host galaxies is at $z \ga 4$.
\end{itemize}

\begin{acknowledgements}
  We thank T. Murayama for useful comments and G. Ferland
  for providing the excellent photoionization code Cloudy
  to the public. 
  TN acknowledges financial support from the Japan Society for the
  Promotion of Science (JSPS) through JSPS Research Fellowship for 
  Young Scientists. RM acknowledges financial support from MIUR
  under grant PRIN-03-02-23.
\end{acknowledgements}

\end{document}